\documentclass[twocolumn,
prl,aps,superscriptaddress
]{revtex4}

\usepackage{graphicx}
\usepackage{amssymb}
\usepackage{amsmath}
\usepackage{amstext}
\usepackage{amsxtra}
\usepackage{gensymb}
\usepackage{bm}
\usepackage[usenames,dvipsnames]{color}
\usepackage{comment}
\usepackage{times}

\newcommand{\vect}[1]{{\mathbf #1}}
\newcommand{\Frac}[2]{\displaystyle\frac{#1}{#2}}

\begin{document}

\title{Non-equilibrium Berezinskii-Kosterlitz-Thouless Transition in a
  Driven Open Quantum System}
	
\author{G.~Dagvadorj} 
\email[]{G.Dagvadorj@warwick.ac.uk} 
\affiliation{Department of Physics, University of Warwick, Coventry,
  CV4 7AL, UK}

\author{J.~M.~Fellows}
\affiliation{Department of Physics, University of Warwick, Coventry,
  CV4 7AL, UK}

\author{S. Matyja\'skiewicz} 
\affiliation{Department of Risk Methodology, Nomura International Plc,
  1 Angel Lane, London, EC4R 3AB, UK}

\author{F.~M.~Marchetti}
\affiliation{Departamento de F\'isica Te\'orica de la Materia
  Condensada \& Condensed Matter Physics Center (IFIMAC), Universidad
  Aut\'onoma de Madrid, Madrid 28049, Spain}

\author{ I.~Carusotto }
\affiliation{INO-CNR BEC Center and Universit\'a di Trento, via
  Sommarive 14, I-38123 Povo, Italy}

\author{M.~H.~Szyma\'nska} 
\email[Corresponding author: ]{M.Szymanska@ucl.ac.uk} 
\affiliation{Department of Physics and Astronomy, University College
  London, Gower Street, London, WC1E 6BT, UK}

\date{\today}



\maketitle

\textbf{ The Berezinskii-Kosterlitz-Thouless mechanism, in which a
  phase transition is mediated by the proliferation of topological
  defects, governs the critical behaviour of a wide range of
  equilibrium two-dimensional systems with a continuous symmetry,
  ranging from superconducting thin films to two-dimensional Bose
  fluids, such as liquid helium and ultracold atoms.
We show here that this phenomenon is not restricted to thermal
equilibrium, rather it survives more generally in a dissipative highly
non-equilibrium system driven into a steady-state.
By considering a light-matter superfluid of polaritons, in the
so-called optical parametric oscillator regime, we demonstrate that it
indeed undergoes a vortex binding-unbinding phase
transition. Yet, the exponent of the power-law decay of the first
  order correlation function in the (algebraically) ordered phase can
  exceed the equilibrium upper limit -- a surprising occurrence, which
  has also been observed in a recent experiment
  \cite{Roumpos2012a}. Thus we demonstrate that the ordered phase is
  somehow more robust against the quantum fluctuations of
  driven systems than thermal ones in equilibrium.}

The Hohenberg-Mermin-Wagner theorem prohibits spontaneous symmetry
breaking of continuous symmetries and associated off-diagonal
long-range order for systems with short-range interactions at thermal
equilibrium in two (or fewer) dimensions \cite{Mermin1966}.
This is because long-range fluctuations due to the soft Goldstone mode
are so strong as to be able to ``shake apart'' any possible
long-ranged order.
The Berezinskii-Kosterlitz-Thouless (BKT) mechanism (for an
  overview see Refs.~\cite{Chaikin2000,Minnhagen1987}) provides a
loophole to the Hohenberg-Mermin-Wagner theorem: Two-dimensional (2D)
systems can still exhibit a phase transition between a
quasi-long-range ordered phase below a critical temperature, where
correlations decay algebraically and topological defects are bound
together, and a disordered phase above such a temperature, where
defects unbind and proliferate, causing exponential decay of correlations. 
Further, it can be shown \cite{Nelson1977} that the algebraic decay
exponent in the ordered phase cannot exceed the upper bound value of $1/4$.

The BKT transition is relevant for a wide class of systems, perhaps
the most celebrated examples are those in the context of 2D
superfluids, as in ${}^4$He and ultracold atoms: Here, despite the
absence of true long-range order, as well as a true condensate
fraction, clear evidence of superfluid behaviour has been observed in
the ordered phase~\cite{Bishop1978}.
Particularly interesting, and far from obvious, is the case of
  harmonically trapped ultracold atomic
  gases~\cite{Hadzibabic2006}. While, in an ideal gas, trapping
  modifies the density of states to allow Bose-Einstein condensation
  and a true condensate~\cite{Shlyapnikov2004,Pitaevskii2003}, weak
  interactions change the phase transition from normal-to-BEC to
  normal-to-superfluid and recover the BKT physics despite the
  system's harmonic confinement.

These considerations are applicable to equilibrium, where the BKT
  transition can be understood in terms of the free energy being
  minimised in either the phase with free vortices or in the one with
  bound vortex-anti-vortex pairs.
However, in recent years a new class of 2D quantum systems has
emerged: strongly driven and highly dissipative interacting many-body
light-matter systems such as, for example, polaritons in semiconductor
microcavities \cite{Carusotto2012}, cold atoms in optical
cavities \cite{Ritsch2013} or cavity arrays
\cite{Carusotto2009,Hartmann2008}.
Due to the dissipative nature of the photonic part, a strong drive is
necessary to sustain a non-equilibrium steady-state.
In spite of this, a transition from a normal to a superfluid
  phase in driven microcavity polaritons has been observed
  \cite{Kasprzak2006,Deng2010}, and the superfluid
  properties of the ordered phase have started being explored
  \cite{Amo2009, Amo2009a,Sanvitto2010,Marchetti2010,Tosi2011}. 
Being strongly driven, the system does not obey the principle of free
energy minimisation, and so it is not obvious whether the transition
between the normal and superfluid phases, as the density of particles
is increased, is of the BKT type i.e. due to vortex-antivortex pairs
unbinding.

Current experiments are not yet able to resolve single shot
  measurements, and so are not sufficiently sensitive to detect
randomly moving vortices.
Algebraic decay of correlations was reported from averaged data
\cite{Roumpos2012a, Roumpos2012}, however, the power-law decay
  displays a larger exponent than is possible in equilibrium, which
  posed questions as to the actual mechanism of the transition. On
the theory side, by mapping the complex Ginzburg-Landau equation
describing long-wavelength condensate dynamics onto the anisotropic
Kardar-Parisi-Zhang (KPZ) equation, Altman et al. \cite{Altman2013}
concluded that although no algebraic order is possible in a truly
infinite system, the KPZ length scale is certainly much larger than
any reasonable system size in the case of microcavity polaritons.

In this work, we consider the case of microcavity polaritons
coherently driven into the optical parametric oscillator
regime~\cite{Stevenson2000, Baumberg2000} as the archetype of a
2D driven-dissipative system.
Another popular pumping scheme is incoherent injection of hot
carriers, which relax down to the polariton ground state by 
exciton formation and interactions with the
lattice phonons \cite{Kasprzak2006,Deng2010}.
However, the incoherently pumped polariton system is challenging
  to model due to the complicated and not yet fully understood
  processes of pumping and relaxation. As a result, one is typically
  forced to use phenomenological models \cite{Chiocchetta2014}, which
  often suffer from spurious divergences. From this point of view, the
  parametric pumping scheme is particularly appealing, as an ab initio
  theoretical description can be developed in terms of a system
  Hamiltonian \cite{Vogel1989}, and its predictions can be directly
  compared to experiment.

Analysing the non-equilibrium steady-state, we show that despite the
presence of a strong drive and dissipation the transition from the
normal to the superfluid phase in this light-matter interacting system
is of the BKT type i.e governed by binding and dissociation of
vortex-antivortex pairs as a function of particle density, and bares a
lot of similarities to the equilibrium counterpart.
However, as recent experiments suggested \cite{Roumpos2012a}, we
  find that larger exponents of the power-law decay are possible
  before vortices unbind and destroy the quasi-long-range order
  leading to exponential decay. This suggests that the external
drive, decay and associated noise favours excitations of collective
excitations, the Goldstone phase modes, which lead to faster spatial
decay, over unpaired vortices which would destroy the
quasi-order all together. This externally over-shaken but not stirred
quantum fluid constitutes an interesting new laboratory to explore
non-equilibrium phases of matter.

\begin{figure}
\includegraphics[width=1\linewidth]{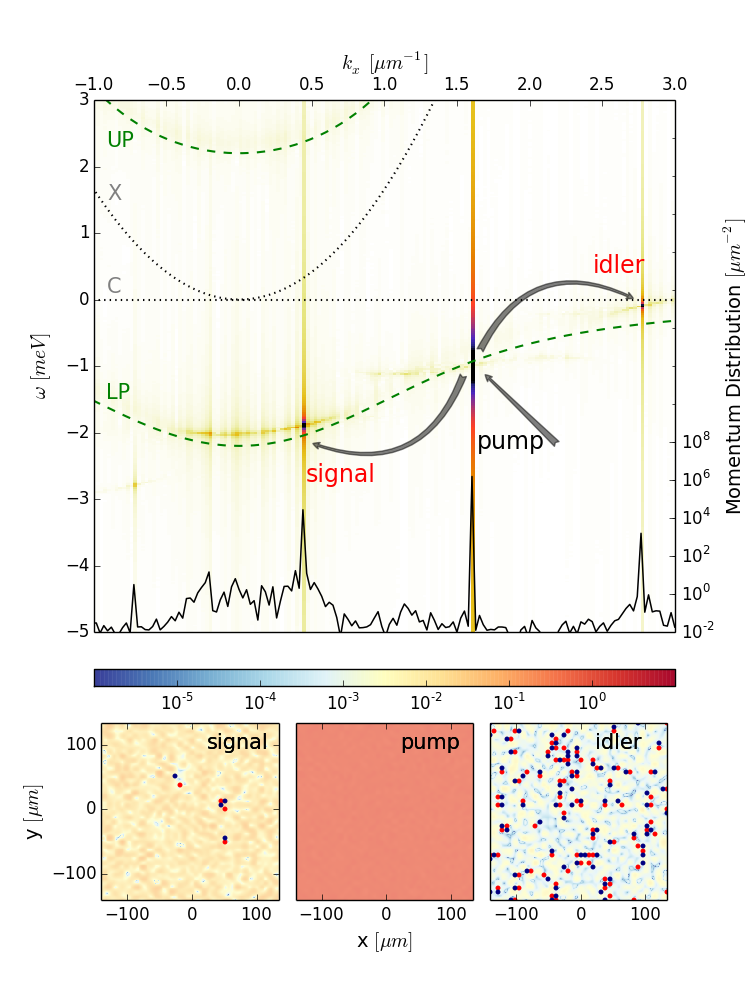}
\caption{\textbf{Polariton system in the OPO regime.}  Upper panel: 2D
  map of the photonic OPO spectrum $|\psi_{C, k_x, k_y=0 }^{}
  (\omega)|^2$ (logarithmic scale) of energy $\omega$ versus the $k_x$
  momentum component (cut at $k_y=0$) for a single noise realisation
  and at a pump power $f_p=1.02436 f_p^{\text{th}}$, where
  $f_p^{^\textrm{th}}$ is the mean-field OPO threshold. The arrows
  show schematically the parametric process scattering polaritons from
  the pump state into the signal and idler modes. Dashed (green) lines
  show the bare upper (UP) and lower polariton (LP) dispersions, while
  dotted (black) lines are the cavity photon ($C$) and exciton ($X$)
  dispersions. The solid (black) line underneath the spectrum is the
  $k_y=0$ cut of the single-shot time-averaged in the steady-state
  momentum distribution $\int dt |\psi_{C, k_x, k_y=0}^{} (t)|^2$,
  clearly showing the macroscopic occupation of the three OPO pump,
  signal, and idler states. Lower panels: 2D maps of the filtered
  space profiles $|\psi_{s,p,i} (\vect{r},t)|^2$ at a fixed time $t$
  for which a steady-state regime is reached --- the pump emission
  intensity is rescaled to $1$. Blue (red) dots indicate the vortex
  (antivortex) core positions.}
\label{fig:figu1}
\end{figure}


\section*{Simulating driven-dissipative open systems}

We describe the dynamics of polaritons in the OPO regime, including the
effects of fluctuations, by starting from the system Hamiltonian for
the coupled exciton and cavity photon field operators
$\hat{\psi}_{X,C}^{} (\vect{r},t)$, depending on time $t$ and 2D
spatial coordinates $\vect{r}=(x,y)$ ($\hbar=1$):
\begin{equation*}
  \hat{H}_S = \int d\vect{r} \begin{pmatrix} \hat{\psi}_{X}^{\dag} &
    \hat{\psi}_{C}^{\dag} \end{pmatrix} \begin{pmatrix}
    \frac{-\nabla^2}{2m_X} + \frac{g_X}{2} |\hat{\psi}_{X}|^2 &
    \frac{\Omega_R}{2} \\ \frac{\Omega_R}{2} &
    \frac{-\nabla^2}{2m_C} \end{pmatrix} \begin{pmatrix}
    \hat{\psi}_{X}^{} \\ \hat{\psi}_{C}^{} \end{pmatrix}\; .
\end{equation*}
Here, $m_{X,C}$ are the exciton and photon masses, $g_X$ the
exciton-exciton interaction strength, and $\Omega_R$ the Rabi
splitting~\cite{Carusotto2012}.
In order to introduce the effects of both an external drive (pump) and
the incoherent decay, we add to $\hat{H}_S$ a system-bath Hamiltonian
$\hat{H}_{SB}$
\begin{multline*} 
  \hat{H}_{SB} = \int d\vect{r} \left[F(\vect{r},t)
    \hat{\psi}_{C}^{\dag} (\vect{r},t) +
    \text{H.c.}\right] \\
  + \sum_{\vect{k}} \sum_{l=X,C} \left\{\zeta^{l}_{\vect{k}}
    \left[\hat{\psi}_{l, \vect{k}}^{\dag} (t) \hat{B}_{l, \vect{k}}^{}
      + \text{H.c.}\right] + \omega_{l, \vect{k}} \hat{B}_{l,
      \vect{k}}^{\dag} \hat{B}_{l, \vect{k}}^{} \right\}\; ,
\end{multline*}
where $\hat{\psi}_{l, \vect{k}}^{} (t)$ are obtained Fourier
transforming to momentum space $\vect{k}=(k_x, k_y)$ the corresponding
field operators in real space $\hat{\psi}_{l}^{}
(\vect{r},t)$. $\hat{B}_{l, \vect{k}}^{}$ and $\hat{B}_{l,
  \vect{k}}^{\dag}$ are the bath's bosonic annihilation and creation
operators with momentum $\vect{k}$ and energy $\omega_{l, \vect{k}}$,
describing the decay processes for both excitons and cavity
photons. To compensate the decay, the system is driven by an external
homogeneous coherent pump $F(\vect{r},t) = f_p e^{i (\vect{k}_p \cdot
  \vect{r} - \omega_p t)}$, which continuously injects polaritons into
a pump state, with momentum $\vect{k}_p$ and energy $\omega_p$.

Within the Markovian bath regime, standard quantum optical
  methods \cite{Szymanska2007,Walls2007} can be used to eliminate the
  environment and obtain a description of the system dynamics in terms
  of a master equation. As the full quantum problem is, in practice,
  intractable, a simple yet useful description of the parametric
  oscillation process properties is provided by the mean-field
approximation, where the quantum fields $\hat{\psi}_{l}^{}
(\vect{r},t)$ are replaced by the classical fields $\psi_{l}^{}
(\vect{r},t)$, whose dynamics is governed by the following generalised
Gross-Pitaevskii equation~\cite{Carusotto2012}
\begin{align}
\label{eq:mfdyn}
  i\partial_t \begin{pmatrix} \psi_{X}^{} \\ \psi_{C}^{} \end{pmatrix}
  &= H_{MF} \begin{pmatrix} \psi_{X}^{} \\ \psi_{C}^{} \end{pmatrix}
  + \begin{pmatrix} 0 \\ F(\vect{r},t) \end{pmatrix}\\ H_{MF}
  &= \begin{pmatrix} \frac{-\nabla^2}{2m_X} + g_X |\psi_{X}|^2 -
    i\kappa_X & \frac{\Omega_R}{2} \\ \frac{\Omega_R}{2} &
    \frac{-\nabla^2}{2m_C} -i \kappa_C\end{pmatrix} \; ,
\nonumber
\end{align}
where $\kappa_{X,C}$ are the exciton and photon decay rates.
By solving~\eqref{eq:mfdyn} both analytically and numerically, much
work has been carried out on the mean-field dynamics for polaritons in
the OPO regime and its properties analysed in
detail~\cite{whittaker05,wouters07:prb,marchetti_review}. Here,
polaritons resonantly injected into the pump state, with momentum
$\vect{k}_p$ and energy $\omega_p$, undergo parametric scattering into
the signal $(\vect{k}_s, \omega_s)$ and idler $(\vect{k}_i, \omega_i)$
states --- see Fig.~\ref{fig:figu1}.
As explained in detail in Ref.~\cite{SM}, as well as in other
works~\cite{marchetti_review}, the full steady-state OPO photon
emission $\psi_{C}^{} (\vect{r},t)$ is filtered in momentum around the
values of the signal, pump and idler momenta $\vect{k}_{s,p,i}$ in
order to get their corresponding steady-state profiles, i.e.,
$\psi_{s,p,i} (\vect{r},t) = \sum_{|\vect{k} - \vect{k}_{s,p,i}| <
  \tilde{k}_{s,p,i}} \psi_{C, \vect{k}}^{} (t) e^{i \vect{k} \cdot
  \vect{r}}$. The choice of each state filtering radius,
$\tilde{k}_{s,p,i}$, is such that the filtered profiles $\psi_{s,p,i}
(\vect{r},t)$ are not affected by them --- for details, see~\cite{SM}.
The mean-field onset of OPO is shown in the inset of
Fig.~\ref{fig:figu3}, where the mean-field densities of both pump and
signal are plotted as a function of the increasing pump power $f_p$.
At mean-field level, parametric processes lock the sum of the phases
of signal and idler fields $\psi_{s,i}$ to that of the external
pump, while allowing a global $U(1)$ gauge symmetry for their phase
difference to be spontaneously broken into the OPO phase --- a feature
which implies the appearance of a Goldstone mode~\cite{Wouters2007}.
As shown below, fluctuations above mean-field can lift, close to the
OPO threshold, this perfect phase locking.

Fluctuations beyond the Gross-Pitaevskii mean-field
description~\eqref{eq:mfdyn} can be included by making use of
phase-space techniques --- for a general introduction, see
Ref.~\cite{Gardiner2004} , while for recent developments in
quantum fluids of atoms and photons, see, e.g.,
Refs.~\cite{Carusotto2005,Giorgetti2007,Foster2010}.
Here, the quantum fields $\hat{\psi}_{l}^{} (\vect{r},t)$ are
represented as quasiprobability distribution functions in the
functional space of C-number fields $\psi_{l}^{} (\vect{r},t)$.
Under suitable conditions, the Fokker-Planck partial differential
  equation, which governs the time evolution of the quasiprobability
  distribution, can be mapped on a stochastic partial differential
  equation, which in turn can be numerically simulated on a finite $N
  \times N$ grid with spacing $a$ (along both $x$ and $y$ directions)
  and a total size $L_{x,y}=Na$ comparable to the polariton pump
    spot size in state-of-the-art experiments. For the system under
  consideration here, the Wigner representation -- one of the many
  possible quasiprobability distributions -- is the most suitable to
  numerical implementation: in the limit $g_X/(\kappa_{X,C} dV) \ll
  1$, where $dV=a^2$ is the cell area, it appears in fact legitimate
  \cite{Drummond1980,Vogel1989} to truncate the Fokker-Planck
  equation, retaining the second-order derivative term only, thus
  obtaining the following stochastic differential equation:
\begin{equation}
  i d \begin{pmatrix} \psi_X \\ \psi_C \end{pmatrix} =
  \left[H_{MF}' \begin{pmatrix} \psi_X \\ \psi_C \end{pmatrix}
    + \begin{pmatrix} 0 \\ F \end{pmatrix}\right] dt +
  i \begin{pmatrix} \sqrt{\kappa_X} dW_X \\ \sqrt{\kappa_C}
    dW_C \end{pmatrix}\; .
\label{eq:wigne}
\end{equation}
Here, $dW_{l=X,C}$ are complex valued, zero-mean, independent
Weiner noise terms with $\langle
  dW^{*}_l (\vect{r},t) dW_m (\vect{r}',t) \rangle =
  \delta_{\vect{r},\vect{r}'} \delta_{l,m} \frac{dt}{dV}$, and the
operator $H_{MF}'$ coincides with $H_{MF}$ in Eq.~\eqref{eq:mfdyn}
with the replacement $|\psi_X|^2 \mapsto |\psi_X|^2-\frac{1}{dV}$.
The same stochastic equation~\eqref{eq:wigne} can be alternatively
derived applying a Keldysh path integral formalism to the Hamiltonian
$\hat{H}_S + \hat{H}_{SB}$, integrating out the bath fields, and
keeping only the renormalisation group relevant
terms~\cite{Sieberer2013}.
Note that, remarkably, some of the difficulties of the truncated
Wigner method met in the context of equilibrium systems, such as for
cold atoms, are suppressed here by the presence of loss and pump terms,
i.e., the existence of a small parameter $g_X/(\kappa_l dV)$ which
controls the truncation~\cite{Carusotto2012}.
Note, however, that the bound on this truncation parameter involves
  the cell area $dV$ of the numerical grid, that is the UV cut-off of
  the stochastic truncated Wigner equation. For typical OPO
  parameters, it is possible to choose $dV$ small enough to capture
  the physics, but at the same time large enough to keep the UV issues
  under control.

We reconstruct the steady-state Wigner distributions $\psi_{l}^{}
(\vect{r},t)$ by considering a monochromatic homogeneous
continuous-wave pump $F(\vect{r},t) = f_p e^{i (\vect{k}_p \cdot
  \vect{r} - \omega_p t)}$ as before and letting the system evolve to
its steady-state. 
In order to rule out any dependence on the chosen initial conditions,
we have considered four extremely different cases: empty cavity with
random noise initial conditions and adiabatic increase of the external
pump power strength; mean-field condensate initial conditions; either
random or mean-field initial conditions in the presence of an
  unpumped region at the edges of the numerical box, so to model a
  sort of “vortex-antivortex reservoir.
The different initial stage dynamics, and their physical interpretation
for each of these four different initial conditions, are carefully
described in~\cite{SM}; in all four cases we always reach the very
same steady-state regime, i.e., all noise averaged observable
quantities discussed in the following lead to the same result --- this
could not be a priori assumed for a non-linear system.

Below we first analyse results from single noise realisations
(concretely, here, for the case of mean-field initial conditions and
no ``V-AV reservoir'' present), by filtering the photon emission at
the signal, pump and idler momenta as also previously done at
mean-field level. The filtered profiles are again indicated as
$\psi_{s,p,i} (\vect{r},t)$ --- for details on filtering see~\cite{SM}.
Second, we consider a large number of independent  noise
realisations and perform stochastic averages of appropriate field
functions in order to determine the expectation values of the
corresponding symmetrically ordered quantum operators. In particular,
we evaluate the signal first-order correlation function as
\begin{equation}
  g^{(1)} (\vect{r}) = \Frac{\langle \psi_{s}^* (\vect{r} + \vect{R},t)
    \psi_{s}^{} (\vect{R},t) \rangle}{\sqrt{ \langle \psi_{s}^*
      (\vect{R},t) \psi_{s}^{} (\vect{R},t) \rangle \langle \psi_{s}^*
      (\vect{r} + \vect{R},t) \psi_{s}^{} (\vect{r} + \vect{R},t)
      \rangle}}\; ,
\label{eq:corre}
\end{equation}
where the averaging $\langle \dots \rangle$ is taken over both noise
realisations as well as the auxiliary position $\vect{R}$, and where
$t$ is either a fixed time after a steady-state is reached, or we
take additional time average in the steady-state~\cite{SM}.

\subsection{Vortices and densities across the transition}

It is particularly revealing to explore the steady-state profiles,
i.e. $\psi_{s,p,i} (\vect{r},t)$, of the signal, pump and idler
states.  Fig.~\ref{fig:figu1} shows a cut at $k_y=0$ of the OPO
spectrum, $|\psi_{C, k_x, k_y=0 }^{} (\omega)|^2$, determined by
solving Eq.~\eqref{eq:wigne} for $\psi_{X,C}^{} (\vect{r},t)$ to a
steady-state and evaluating the Fourier transforms in both space and
time.
Note, that the logarithmic scale of this 2D map plot (which we employ
to clearly characterise all three OPO states) makes the emission
artificially broad in energy, while in reality this is sharp (as
required by a steady-state regime), as well as it is very narrow in
momentum.
The filtered space profiles $\psi_{s,p,i} (\vect{r},t)$ shown in the
bottom panels of Fig.~\ref{fig:figu1} reveal that while the pump state
is homogeneous and free from defects, vortex-antivortex (V-AV) pairs
are present for both signal and idler states. Note, that while at the
mean-field level the sum of the signal and idler phases is locked to
the one of the pump (and thus a V in the signal implies the presence
of an AV at the same position in the idler), the large
  fluctuations occurring in the vicinity of the OPO threshold make
  this coherent phase-locking mechanism only weakly enforced,
resulting in a different number (and different core locations) of V-AV
pairs in the signal and idler states.
Because the density of photons in the idler state is much lower than
the one at the signal (see, e.g., the photonic momentum distribution
plotted as a solid black line inside the upper panel of
Fig.~\ref{fig:figu1}), while both states experience the same noise
strength, the number of V-AV pairs in the filtered photonic signal
profile is much lower than the number of pairs in the filtered
photonic idler profile.
Phase locking between signal and idler is recovered instead for pump
powers well above the OPO threshold, where long-range coherence over
the entire pumping region is re-established.

\begin{figure}
\includegraphics[width=1\linewidth]{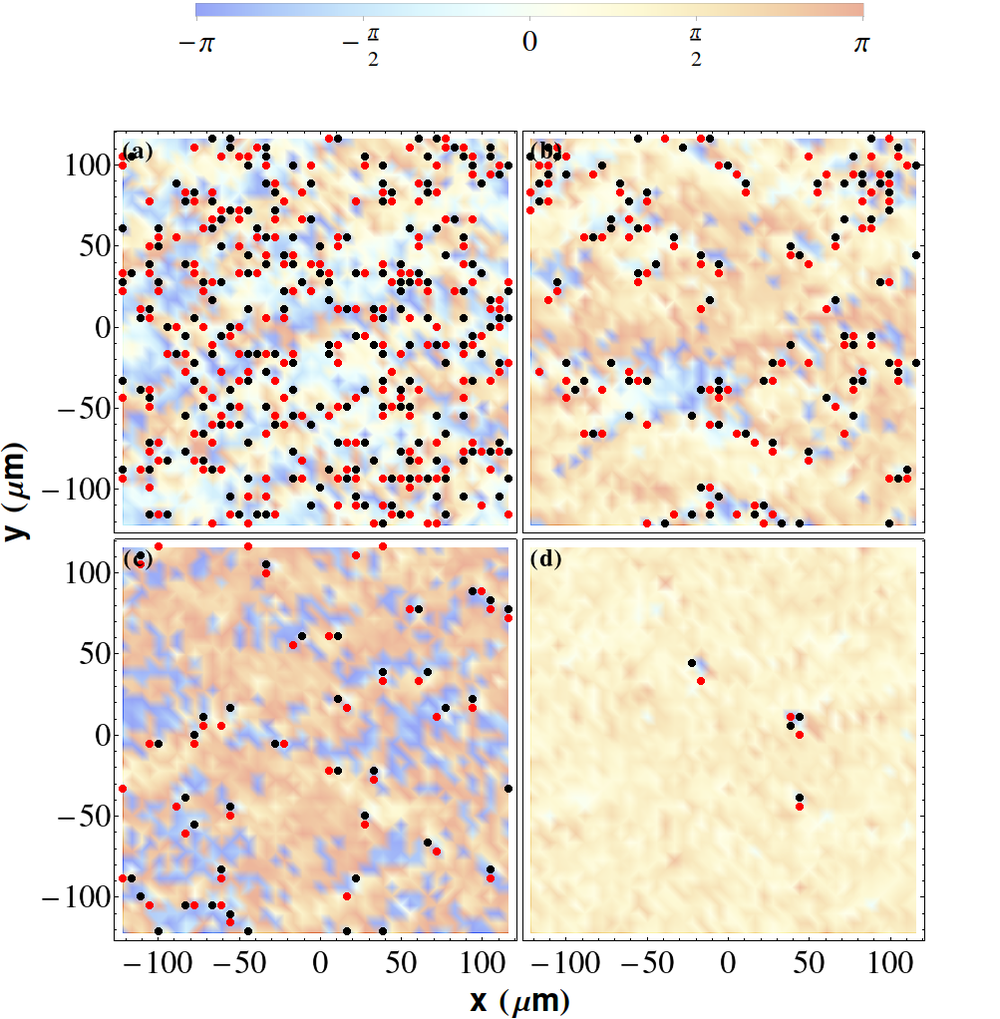}
\caption{\textbf{Binding-unbinding transition and vortex-antivortex
    proliferation across the OPO threshold.} Phase (colour map) of the
  filtered OPO signal $\psi_{s} (\vect{r},t)$ and position of
  vortices (black dots) and antivortices (red dots) for increasing
  values of the pump power, in a narrow region close to the mean-field
  OPO threshold $f_p^{\text{th}}$: (a) $f_p=1.00287f_p^{\textrm{th}}$,
  (b) $f_p=1.01648f_p^{\textrm{th}}$, (c)
  $f_p=1.01719f_p^{\textrm{th}}$, and (d)
  $f_p=1.02436f_p^{\textrm{th}}$. We observe a dramatic decrease of
  both the number of Vs and AVs, as well as the typical distance
  between pairs, as a function of the increasing pump power. The
  filtered profiles are plotted at a late stage of the dynamics, at
  which a steady-state is reached.}
\label{fig:figu2}
\end{figure}
\begin{figure}
\includegraphics[width=1\linewidth]{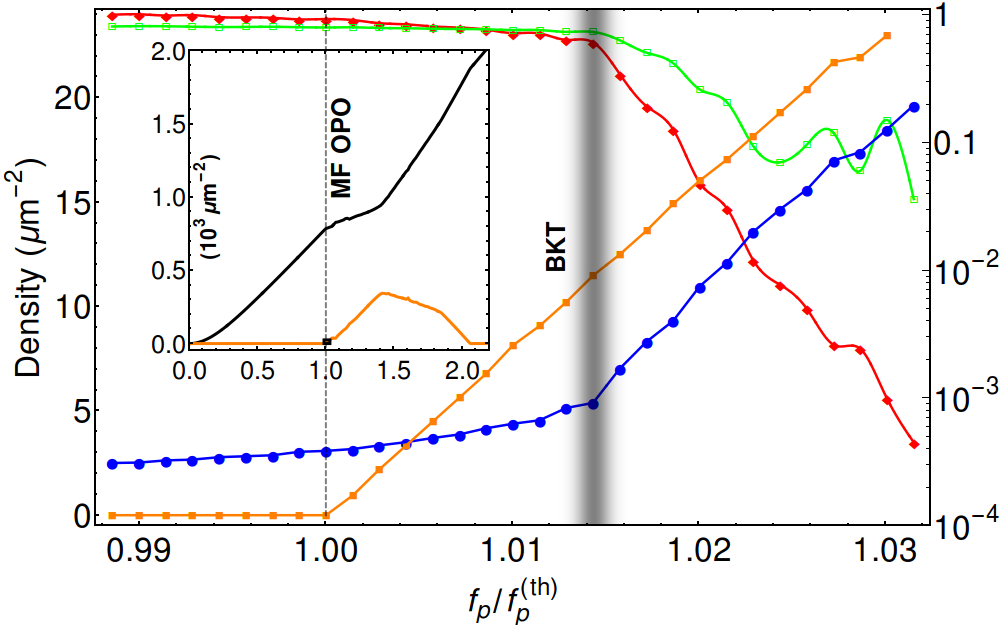} 
\caption{\textbf{The phase diagram and the BKT transition.} Inset:
  Mean-field photonic OPO densities for pump (black) $n_{p}$ and
  signal (orange) $n_{s}$ states as a function of increasing pump
  power $f_p$ rescaled by the threshold value $f_p^{\text{th}}$
  (vertical black dashed line). The black square at $f_p \simeq
  f_p^{\text{th}}$ indicates the tiny pump strength interval close to
  mean-field threshold analysed in the main panel. Main panel: We plot
  with (orange) squares the same mean-field signal density $n_{s}$ as
  in the inset. All other data are noise averaged properties from
  stochastic simulations as a function of the pump strength. The noise
  averaged signal density $n_s$ is plotted with (blue) dots; the
  average vortex number in the signal rescaled by its average maximum
  value, $N_{\textrm{max}}=222.8$ with (red) diamonds; the noise
  averaged and symmetrised distance ratio $\langle b \rangle$ between
  nearest neighbouring V-V and AV-AV over V-AV pairs with (green) empty
  squares.  The shaded region indicates the pump region for
  the BKT transition.}
\label{fig:figu3}
\end{figure}

The proliferation of vortices below the OPO transition, followed
  by a sharp decrease in their density and their binding into close
  vortex-antivortex pairs is illustrated in
Fig.~\ref{fig:figu2}. Here, we plot the 2D maps of the phase for the
single noise realisation of the filtered OPO signal $\psi_{s}
(\vect{r},t)$ (photonic component) for increasing values of the pump
power $f_p$ in a narrow region close to the mean-field OPO threshold
$f_p^{\text{th}}$; the position of the generated vortices
(antivortices) are marked with blue (black) dots.
While at lower pump powers there is a dense ``plasma'' of Vs and
AVs, the number of V-AV pairs decrease with increasing pump powers
till eventually disappearing altogether (not shown). We do also record
a net decrease in the distance between nearest neighbouring vortices
with opposite winding number with respect to that between vortices
with the same winding number. In order to quantify the vortex binding
across the OPO transition, we measure, for each detected vortex, the
distance to its nearest vortex, $r_{\text{V-V}}$ and to its nearest
antivortex $r_{\text{V-AV}}$; and similarly, for each detected
antivortex, we measure $r_{\text{AV-AV}}$ and $r_{\text{AV-V}}$. We
then consider the symmetrised ratio $b =
\frac{r_{\text{V-V}}+r_{\text{AV-AV}}}{r_{\text{V-AV}}+r_{\text{AV-V}}}$. In
order to extract a noise realisation independent quantity, an average
over many different realisations, as well as over
individual vortex positions, is performed to obtain $\langle b
\rangle$; this quantity $\langle b \rangle \to 1$ for an unbound
vortex plasma, while $\langle b \rangle \to 0$ when vortices form
tightly bound pairs.
We observe a dramatic drop in $\langle b \rangle$ (green squares in
Fig.~\ref{fig:figu3}) when increasing the pump power across the OPO
threshold, indicating that vortices and antivortices are indeed
binding, as it is expected for a BKT transition.

By evaluating other relevant noise averaged observable quantities, we
are able to construct a phase diagram for the OPO transition in
Fig.~\ref{fig:figu3} and link it with the properties of the BKT
transition.
We evaluate the averaged signal photonic density at some time $t$
  in the steady-state, $n_{s} = \int d\vect{r} \langle |\psi_{s}
  (\vect{r},t)|^2 \rangle /V$, where $V=(Na)^2$ is the system area and
  $\langle \dots \rangle$ indicates the noise average for the
  stochastic dynamics (blue dots). We also show the steady state
  signal density in the mean field (orange squares). The corresponding
  mean-field densities for both signal (orange line) and pump (black
  line) are presented for comparison in the inset of
  Fig.~\ref{fig:figu3}.
At mean-field level, both signal and idler (not shown) suddenly switch
on at the OPO threshold pump power, $f_p = f_p^{\text{th}}$ and both
states are macroscopically occupied above threshold.
The effect of fluctuations is to smoothen the sharp mean-field
transition, as clearly shown by the (blue) dots in the main panel of
Fig.~\ref{fig:figu3}, where we plot the noise average signal density
$n_{s}$. This is because, even below the mean-field threshold,
incoherent fluctuations weakly populate the signal. Note also that,
even though somewhat smoothened, we can still appreciate a kink in the
$n_{s}$ density, but at higher values of the pump power compared to
the mean-field threshold $f_p^{\text{th}}$. We identify this as the
novel BKT transition for our out-of-equilibrium system, as discussed
more in detail below.
This is further confirmed by a sudden decrease of the averaged number
of vortices in the signal (red diamonds), and of the averaged
distance between nearest neighbouring vortices of opposite winding
number, $\langle b \rangle$ (green squares), as a function of
the pump power concomitant with the observed kink for $n_{s}$.
These results suggest that the system undergoes an OPO transition
which, by including fluctuations above mean-field, is indeed analogous
to the equilibrium BKT transition. Both vortices and antivortices
proliferate below some threshold and, above, they bind to eventually
disappear altogether. As indicated by the black square in the inset of
Fig.~\ref{fig:figu3}, the region for such a crossover is indeed narrow
in the pump strength.

\begin{figure}
\includegraphics[width=1\linewidth]{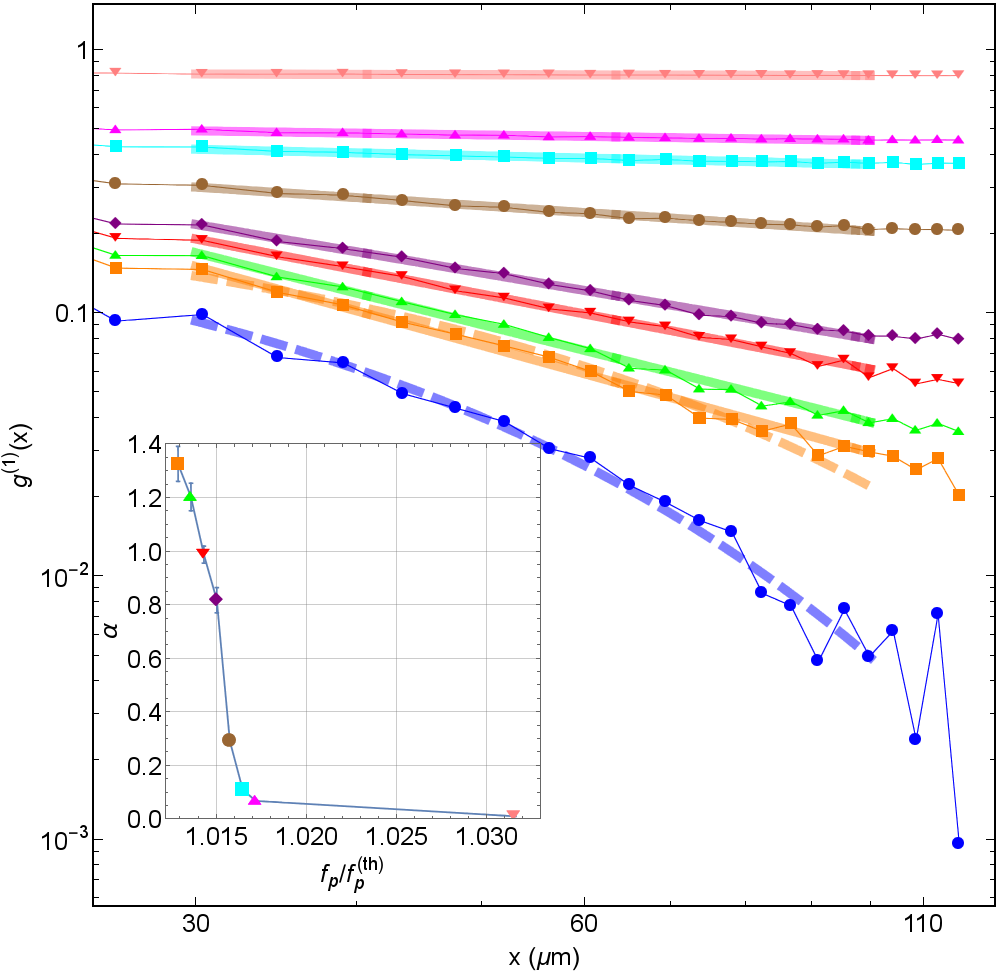}
\caption{\textbf{Algebraic and exponential decay of the first order
    correlation function across the BKT transition.} Main panel:
  Long-range spatial dependence of $g^{(1)} (\vect{r})$ for different
  pump powers $f_p/f_p^{\text{th}}$ close to the mean-field pump
  threshold (the symbols are the same ones as in the inset and
  correspond to the same values of $f_p/f_p^{\text{th}}$). Thick solid
  (thick dashed) lines are power-law (exponential) fitting, from which
  values of the exponent $\alpha$  are
  derived. The $f_p/f^\textrm{th}_p=1.0129$ case (orange squares) is
  a marginal case where both algebraic and exponential fits apply almost
  equally well, signalling the BKT transition region. Inset: Power-law
  algebraic decay exponent $\alpha$ for different pump powers
  $f_p/f_p^{\text{th}}$; error bars are standard deviations of the
  time-average.}
\label{fig:figu4}
\end{figure}
%


\section*{First-order spatial correlations}
For systems in thermal equilibrium, the BKT transition is associated
with the onset of quasi-off-diagonal long-range order, i.e., with the
algebraic decay of the first-order correlation function in the ordered
phase, where vortices are bound, and exponential decay in the
disordered phase, where free vortices do proliferate. In order to
investigate whether the same physics applies to our out-of-equilibrium
open-dissipative polariton system, we evaluate the signal first-order
correlation function $g^{(1)} (\vect{r})$ according to the
prescription of Eq.~\eqref{eq:corre} and characterise its long-range
behaviour in Fig.~\ref{fig:figu4}.
We observe the ordering transition as a crossover in the long-distance
behaviour between an exponential decay in the disordered phase,
$g^{(1)}(\vect{r}) \sim e^{-r/\xi}$, and an algebraic decay in the
quasi-ordered phase, $g^{(1)} (\vect{r}) \sim (r/r_0)^{-\alpha}$.  We
therefore fit the tail of the calculated correlation function to both
of these functional forms and observe that, at the onset of vortex
binding-unbinding and proliferation, the signal's spatial correlation
function changes its long-range nature, from exponential at lower pump
powers to algebraic at higher (see Fig.~\ref{fig:figu4}).

However, in contrast with the thermal equilibrium case, we do observe
that the exponent $\alpha$ of the power-law decay (inset of
Fig.~\ref{fig:figu4}) can exceed the equilibrium upper bound of $1/4$
\cite{Nelson1977}, and can reach values as high as $\alpha \simeq
1.2$ for $f_p/f_p^{\text{th}} = 1.0136$, just within the ordered
phase.
Further, as thoroughly discussed in Ref.~\cite{SM}, it is interesting
to note that, close to the transition, we do observe a critical
slowing down of the dynamics: Here, the convergence to a steady-state
is dramatically slowed down compared to cases above or below the OPO
transition, a common feature of other phase transitions. At
the same time, close to threshold, the convergence of noise averaged
number of vortices is much faster than the convergence of the
power-law exponent $\alpha$. This indicates that the fluctuations
induced by both the external drive and the decay preferentially excite
collective excitations, rather than topological excitations, resulting
in vortices being much more dynamically stable.
Finally, note that for sufficiently strong pump powers, the power-law
exponent becomes extremely small and thus quasi-long-range order is
difficult to distinguish from the true long-range order over the
entire system size.

Our findings explain why recent experimental studies, both in the OPO
regime~\cite{Spano2013}, as well as for non-resonant
pumping~\cite{Roumpos2012a}, experienced noticeable difficulties in
investigating the power-law decay of the first order correlation
function across the transition.
We do indeed find that the pump strength interval over which power-law
decay can be clearly observed is extremely small, and the system
quickly enters a regime where coherence extends over the entire system
size, as measured in~\cite{Spano2013}.
This was also observed in non-resonantly pumped experiments when using
a single-mode laser \cite{Kasprzak2006}. However, by intentionally
  adding extra fluctuations with a multimode laser pump, as
in~\cite{Roumpos2012a}, power-law decay was finally observed in the
correlated regime, with an exponent in the range $\alpha \simeq
0.9-1.2$, in agreement with our results.


\section*{Discussion}

Using microcavity polaritons in the optical parametric oscillator
  regime as the prototype of a driven-dissipative system, we have
  numerically shown that a mechanism analogous to the BKT transition,
  which governs the equilibrium continuous-symmetry-breaking phase
  transitions in two dimensions, occurs out of equilibrium for a
  driven-dissipative system of experimentally realistic size.
Notwithstanding the novelty and significance of this result, there are
a number of novel features which warrant discussion as they are
peculiar to non-equilibrium phase transitions.
We have observed that the exponent of algebraic decay in the
quasi-long-range ordered phase exceeds what would be attainable in
equilibrium. This recovers a recent observation \cite{Roumpos2012a},
and strongly suggests that indeed a non-equilibrium BKT may have been
seen there.
Moreover, our findings imply that the ordered phase is more robust to
fluctuations induced by the external drive and decay than an analogous
equilibrium ordered phase would be to thermal fluctuations.
Although for realistic experimental conditions, we have found that the
region for BKT physics, before the pump power is strong enough to
induce perfect spatial coherence over the entire system size, is
indeed narrow, we believe our work will encourage further experimental
investigations in the direction of studying the non-equilibrium BKT
phenomena.
Even though the small size of the critical region has so far
  hindered its direct experimental study, our calculations indicate
  that the macroscopic coherence observed in past polariton
  experiments
\cite{Keeling2007, Kasprzak2006, Stevenson2000, Baumberg2000, Carusotto2012, Deng2010, Spano2013}
results from a non-equilibrium phase transition of the BKT rather
  than the BEC kind.


\section*{Methods}

We simulate the dynamics of the stochastic equations~\eqref{eq:wigne}
with the XMDS2 software framework~\cite{Dennis2012} using a fixed-step 
(where the fixed step-size ensures stochastic noise consistency) 
4th order Runge-Kutta (RK) algorithm, which we have tested against fixed-step 
9th order RK, and a semi-implicit fixed-step algorithm with 3 and 5 iterations.
We choose the system parameters to be close to current
experiments~\cite{Sanvitto2010}: The Rabi frequency is chosen as
$\Omega_R=4.4$~meV, the mass of the microcavity photons is taken
  to be $m_C=2.3\times10^{-5}m_e$, where $m_e$ is the electron mass,
  the mass of the excitons is much greater than this so we may take
  $m_X^{-1}\to0$, the exciton and photon decay rates as $\kappa_X =
\kappa_C = 0.1$~meV, and the exciton-exciton interaction strength
$g_X=0.002$~meV$\mu$m$^2$ \cite{Ferrier2011}.
The pump momentum $\vect{k}_p = (k_p,0)$, with $k_p =
1.6$~$\mu$m$^{-1}$, is fixed just above the inflection point of the LP
dispersion, and its frequency, $\omega_p - \omega_X (0)=1.0$~meV, just
below the bare LP dispersion.
In order to satisfy the condition necessary to derive the truncated
Wigner equation~\eqref{eq:wigne}, $g_X/(\kappa_{X,C} d V) \ll 1$,
whilst maintaining a sufficient spatial resolution and, at the same
time, a large enough momentum range so that to resolve the idler
state, simulations are performed on a 2D finite grid of $N \times N =
280 \times 280$ points and lattice spacing $a = 0.866$~$\mu$m.
Thus, the only system parameter left free to be varied is the pump
strength $f_p$: We first solve the mean-field
dynamics~\eqref{eq:mfdyn} in order to determine the pump threshold
$f_p^{\text{th}}$ for the onset of OPO. We then vary the value of
$f_p$ around $f_p^{\text{th}}$ in presence of the noise in order to
investigate the nature of the OPO transition.
We analyse the results from single noise realisations by filtering
the full photonic emission for signal pump and idler, as described in
the main text, as well as in~\cite{SM}.
Further, we average all of our results over many independent
  realisations, which are either taken from $96$ independent stochastic
  paths or from multiple independent snapshots in time after the
steady-state is reached: As thoroughly discussed in~\cite{SM}, 96
  stochastic paths is shown to be sufficient to ensure the
convergence of noise averaged observable quantities.
For each noise realisation, vortices are counted by summing the phase
difference (modulo $2\pi$) along each link around every elementary
plaquette on the filtered grid. In the absence of a topological defect
this sum is zero, while if the sum is $2\pi$ ($-2\pi$) we determine
there to be a vortex (antivortex) at the center of the plaquette.
The number of vortices is then averaged over the different stochastic
paths or over time in the steady-state (see Ref. \cite{SM}): We
consider the average number of vortices to be converged in time when
its variation is less then $5\%$.
Finally, the first order correlation function $g^{(1)} (\vect{r})$ is
evaluated according to Eq.~\eqref{eq:corre}, by averaging over both
the noise and the auxiliary position $\vect{R}$; as discussed
in~\cite{SM}, this can be computed efficiently in momentum space.





\iftrue

\fi


\

\

\noindent 
\textbf{Acknowledgments}\\
\noindent 
We thank J.~Keeling for stimulating discussions.
MHS acknowledges support from EPSRC (grants EP/I028900/2 and
EP/K003623/2),
FMM from the programs MINECO
(MAT2011-22997) and CAM (S-2009/ESP-1503), and 
IC from ERC through the QGBE grant
and from the Autonomous Province of Trento, partly through the project
``On silicon chip quantum optics for quantum computing and secure
communications'' (``SiQuro'').


\newpage

\setcounter{page}{1}

\pagenumbering{gobble}

\setcounter{secnumdepth}{2}

\renewcommand{\figurename}{\textsc{S.~Fig.}}

\setcounter{equation}{0}
\setcounter{figure}{0}

\onecolumngrid
\begin{center}
\large \textbf{Supplemental Material for ``Non-equilibrium
  Berezinskii-Kosterlitz-Thouless Transition in a Driven Open Quantum
  System''}
\end{center}
\vspace{3ex}
\twocolumngrid

In this supplementary information we provide a more detailed
  account of some of the technical issues arising from our numerical
  methods, which are not of sufficiently general interest to warrant a
  discussion in the main text, but which are germane to the validity of
  our conclusions.
In particular, we discuss the filtering we perform in order to isolate
the signal state, and the loss in resolution this incurs, and we
describe various tests we have undertaken in order to ensure that the
results we present are properly converged.

\section{Momentum filtering}

In order to study the condensate at the signal (or idler), we have to
filter the full emission $\psi_{C}^{} (\vect{r},t)$ in such a way as
to omit contributions with momentum outside a set radius about the
signal (idler) states in $\mathbf{k}-$space.
This applies both to the theory and the experimental data, and
  this procedure could equivalently be performed in either momentum or
  energy space. Here we filter in momentum space,
  and define:
 \begin{equation*} 
\psi_{s,p,i} (\vect{r},t) = \sum_{|\vect{k} - \vect{k}_{s,p,i}| <
  \tilde{k}_{s,p,i}} \psi_{C, \vect{k}}^{} (t) e^{i \vect{k} \cdot
  \vect{r}},
 \end{equation*}
where $\vect{k}_{s,p,i}$ is the momentum of the signal, pump and idler
states respectively and $\tilde{k}_{s,p,i}$ is the filtering radius.
We fix the filtering radius for the signal and idler to be
$\tilde{k}_{s,i}=\tfrac{1}{2}|\mathbf{k}_{s,i}-\mathbf{k}_p|$ and
choose to operate in the frame of reference co-moving with the signal
(idler) so that the (physically irrelevant) background current does
not show in the data.
The filtering reveals the phase-freedom of the signal (idler) state at
the expense of spatial resolution. Note, that this limits our ability
to distinguish vortex-antivortex (V-AV) pairs with a separation
less than a distance $\pi/\tilde{k}_{l}$, which for our chosen
parameters is $\approx2.708\mu\textrm{m}$. However, such extremely
close vortex-antivortex pairs do not affect the spatial correlations
of the field at large distances.

\section{Numerical Convergence}
\subsection{Number of stochastic realisations}

\begin{figure}[h]
\includegraphics[width=1\linewidth]{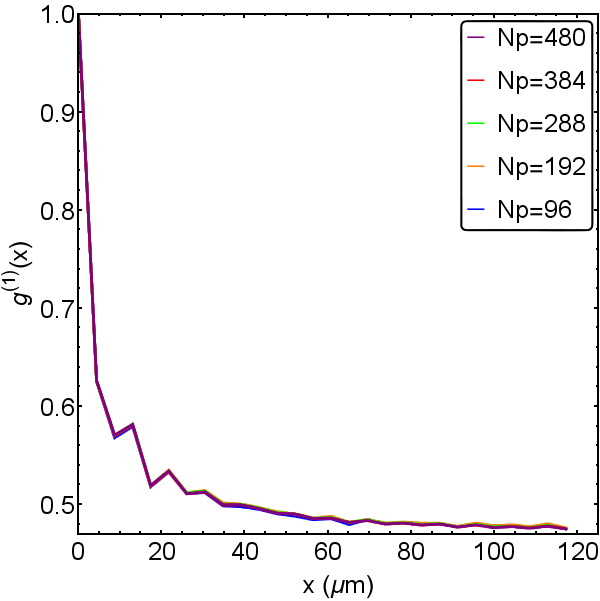}
\caption{ {\bf Correlation function averaged over different numbers of
    realisations.}
The strength of the pump is fixed at $f_p=1.017 f_p^{\textrm{th}}$,
and each simulation is run well into the steady state. We calculate
the correlation function averaged over 96, 192, 288, 384 and 480
realisations and observe no significant improvement in the convergence
past 96 realisations.  }
\label{fig:g1_vs_np}
\end{figure}

The stochastic averages over the configurations of different realisations
of the fields provide the expectation value of the corresponding
symmetrically ordered operators.
The realisation-averaged results presented in the main paper have been
averaged over 96 realisations taken from independent stochastic paths
as we assessed this to be sufficiently many to give a reliably
converged result.  In Fig. \ref{fig:g1_vs_np} we show the first order
correlation function $g^{(1)}(x)$ for $f_p=1.017 f_p^{\textrm{th}}$,
averaged over different numbers of realisations (96, 192, 288, 384 and
480). We conclude that 96 is sufficient to determine the nature of the
correlations.

The average number of vortices also does not change significantly (no
more than $\pm0.5$) beyond 96 realisations, and so we conclude that
this number is sufficient to ensure consistent results in both the
smooth and topological sectors of the model.

\subsection{Convergence in time to a steady state}

\begin{figure}[h]
\includegraphics[width=1\linewidth]{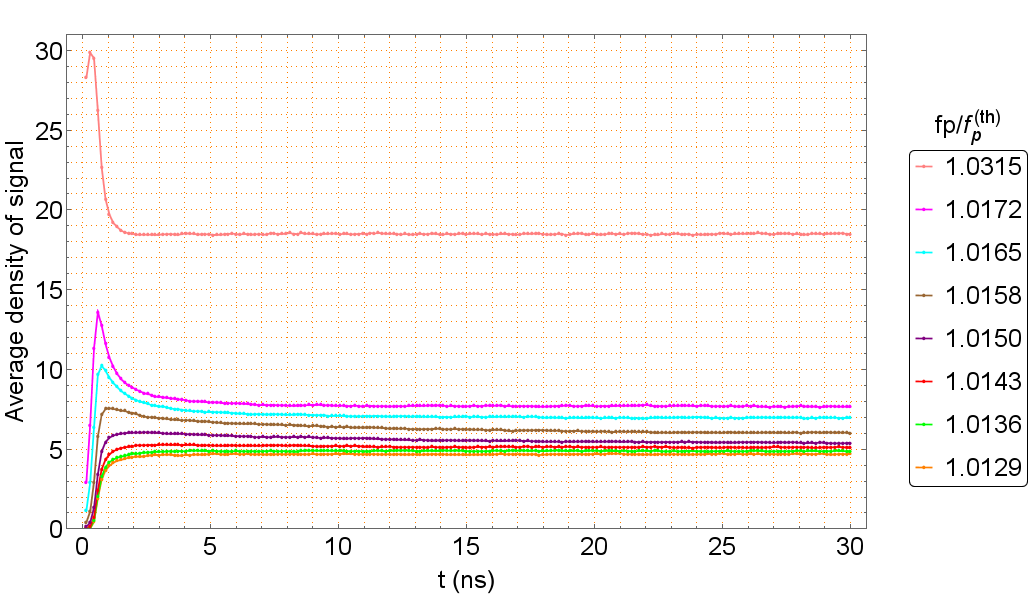}
\caption{ {\bf Convergence in time of the average signal density.}
  Here we show $|\psi_\textrm{s}|^2$ evolving in time for a range of
  pump powers. The initial jump in the signal occupation is induced by
  the introduction of stochastic processes, and dies away as the system
  resolves toward its non-equilibrium steady state.  }
\label{fig:ns_vs_t}
\end{figure}

\begin{figure}[h]
\includegraphics[width=1\linewidth]{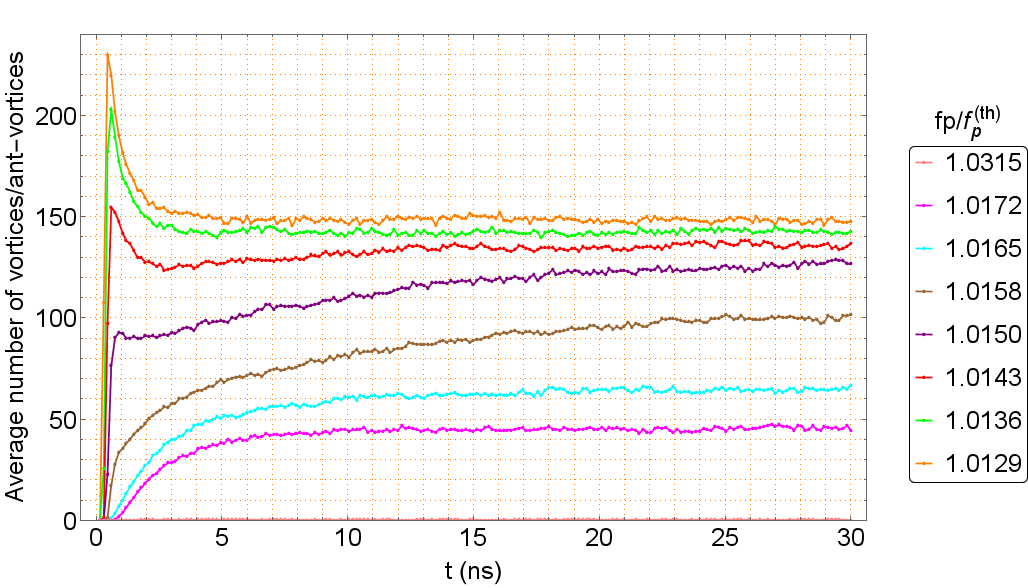}
\caption{ {\bf Convergence in time of the average number of vortices.}
  Here we show the average number of vortices in the system
  evolving in time for a range of pump powers. The introduction of
  stochastic processes increases the number of vortices (which is
  always zero in the spatially homogeneous mean field steady-state),
  which then evolves toward its steady state value through pair
  creation and annihilation events.  }
\label{fig:nv_vs_t}
\end{figure}

In order to assess when a time evolution has reached its steady state we
consider the average (over realisations) of both the signal density,
$|\psi_\textrm{s}|^2$, and the number of vortices. In Figs.
\ref{fig:ns_vs_t} and \ref{fig:nv_vs_t} we show the evolution of these
quantities toward a steady state, assuming the initial condition
wherein the system is allowed to reach its mean field steady state (at
time $t=0$) and stochastic processes are adiabatically switched on.
In practice this means that the noise terms are multiplied by a ramp
function
\[
r(t) = \frac{1}{2}\left[ \tanh\left(\frac{t-t_0}{t_r}\right)+1 \right]
\]
\noindent such that the noise is slowly increases from $0$, achieving
half its maximum at $t_0$, at a rate determined by $t_r$. For the
empty cavity with noise initial condition, in which the pump is
adiabatically increased, it is the pump term that is multiplied by
this ramp function. We choose $t_0 = 450\textrm{ps}$ and $t_r =
150\textrm{ps}$ as these prove to give a fairly rapid convergence to
the steady state. It is, however, important to note that the steady
state is unique and does not depend upon the specific values of $t_0$
and $t_r$ we choose nor on the initial conditions, which we start the
dynamics from.

In all cases the steady state is reached within around twenty
nanoseconds. The slowest convergence occurs in the vicinity of the BKT
transition. This critical slowing of the dynamics is to be anticipated
given the divergence of the correlation length as the transition is
approached.

\subsection{Realisations from independent stochastic paths
  vs. independent time snapshots in a single
  path} 

\begin{figure}[h]
\includegraphics[width=1\linewidth]{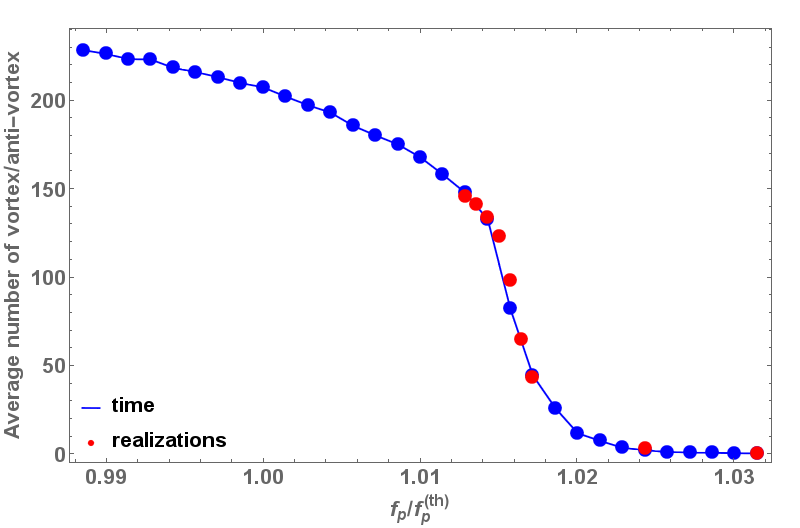}
\caption{ {\bf Time versus stochastic realisations averaging.}
  The red dots show number of vortices averaged over many
    stochastic paths. The blue dots show the number of vortices
    averaged over the final 500 frames, running from $15\textrm{ns}$
    to $30\textrm{ns}$, of a single stochastic path, once the system
    has converged to a steady state.
We observe an excellent agreement between these two approaches and so
away from the critical region we can consider the (less numerically
intensive) average over time.  }
\label{fig:nv_vs_fp}
\end{figure}

In Fig. \ref{fig:nv_vs_fp} we show the number of vortices for a broad
range of pump powers across the transition averaged over realisations
taken from independent stochastic paths (red dots) and from multiple
snapshots over time (blue dots) once the steady-state is reached.
Averaging over time within the steady state is a less numerically
intensive approach, and shows an excellent agreement with the average
over realisations even in the critical region.
In practice, averaging over time is computationally efficient
  away from the critical region, where the steady state is reached
  quickly and memory is quickly lost during the time evolution. Around
  the critical region it takes a significant time to reach a steady
  state and then to decorrelate the snapshots, so averaging over
  stochastic paths becomes more computationally effective.

\subsection{Dependence on different initial conditions}

We have investigated a number of physically diverse initial conditions
in order to rule out dependence of the final steady state on the
starting configuration.  To initiate our simulations we either
adiabatically increase the pump strength atop a white noise background
or adiabatically increase the strength of the stochastic terms
starting from the mean field steady state.
We also wish to eliminate the possible effects of trapped
  vortices, and so in addition to simulations with a uniform pump
  region we have performed simulations, where a small strip around the
  numerical integration box is left un-pumped so as to act as a
  source/sink for vortices.
The four distinct initial conditions we consider are therefore
  those classified in the following table:

\vspace{5pt}\begin{tabular}{c|c|c|} 
& ~Increasing Pump~ & ~Increasing Noise~ \\ 
\hline 
No Reservoir & Scheme A & Scheme B \\ 
\hline 
Reservoir & Scheme D & Scheme C \\ 
\hline 
\end{tabular}\vspace{5pt}

In schemes A and D, there are initially very many vortices, which then
proceed to annihilate with one another so that the overall vorticity
tends to decrease toward the steady state. In schemes B and C there
are no vortices at the outset but as stochastic processes shake the
system the vorticity increases toward the steady state. Therefore even
before the true steady state is reached, we can treat schemes A and D
as upper bounds for the vorticity and schemes B and C as lower bounds.
In schemes C and D we incorporate a reservoir (region with no drive
and thus of very low density) of width $34.72\textrm{nm}$ along
  the two sides of the numerical integration box parallel to $k_p$,
which we then exclude from calculations of the vorticity, signal
density, and correlations. The width of the reservoir does not change
the result except in that it needs to be wide enough that the
condensate has room to decay away to zero.

In Supplementary Movie 1 we show the evolution towards the steady
state for all four of these schemes with a pump power $f_p=1.017
f_p^{\textrm{th}}$. We plot the vortex density (top panel) as well as
dynamics of vortices (bottom panel, where blue and red dots show
positions of V and AV cores) for each scheme (A, B, C and D starting
from the left).
We see that every scheme converges fairly rapidly towards the same
steady state. From the animations it is evident that the vortices
always appear in pairs but that these pairs are not always tightly
bound.  It was not {a priori} obvious that all four schemes should
lead to the same steady state but we take this as evidence that the
physical process leading to the steady state is universal for this
system.
We observe that the schemes incorporating an empty reservoir converge
more quickly to the steady state than their counterparts without a
reservoir. Nevertheless we present data for scheme B because the
reservoirs have no counterpart in the traditional BKT transition to
which we wish to compare our results.

\section{Fitting to the spatial correlation function}

\begin{figure*}[h]
\includegraphics[width=0.49\linewidth]{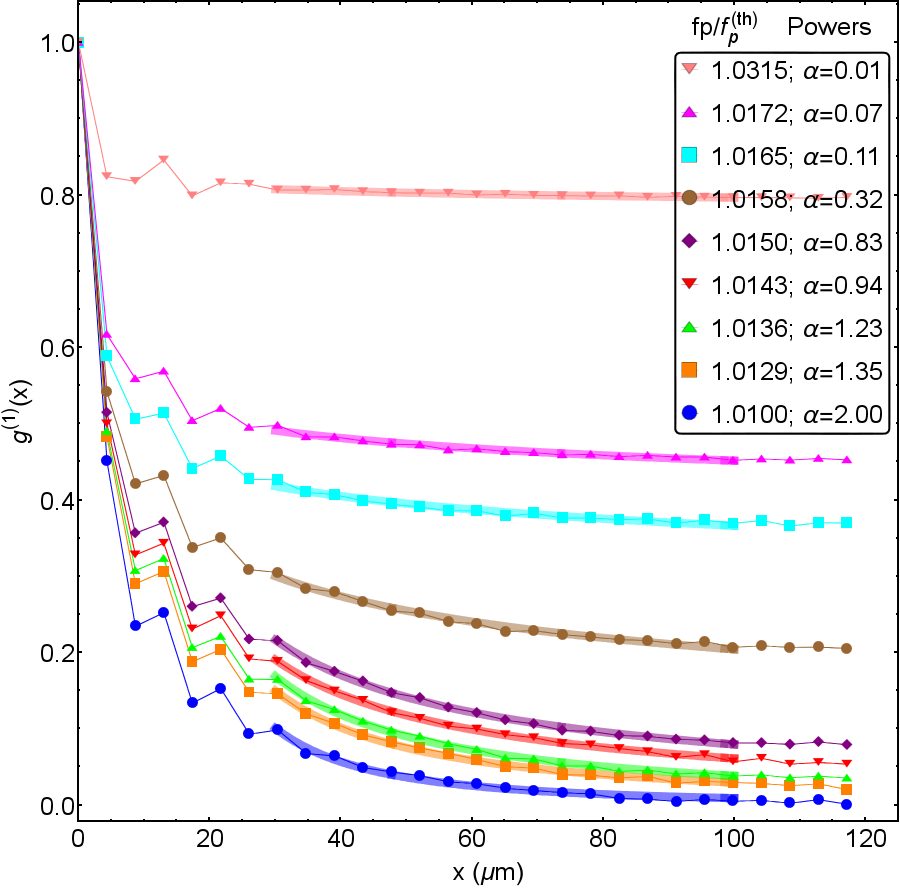}
\includegraphics[width=0.49\linewidth]{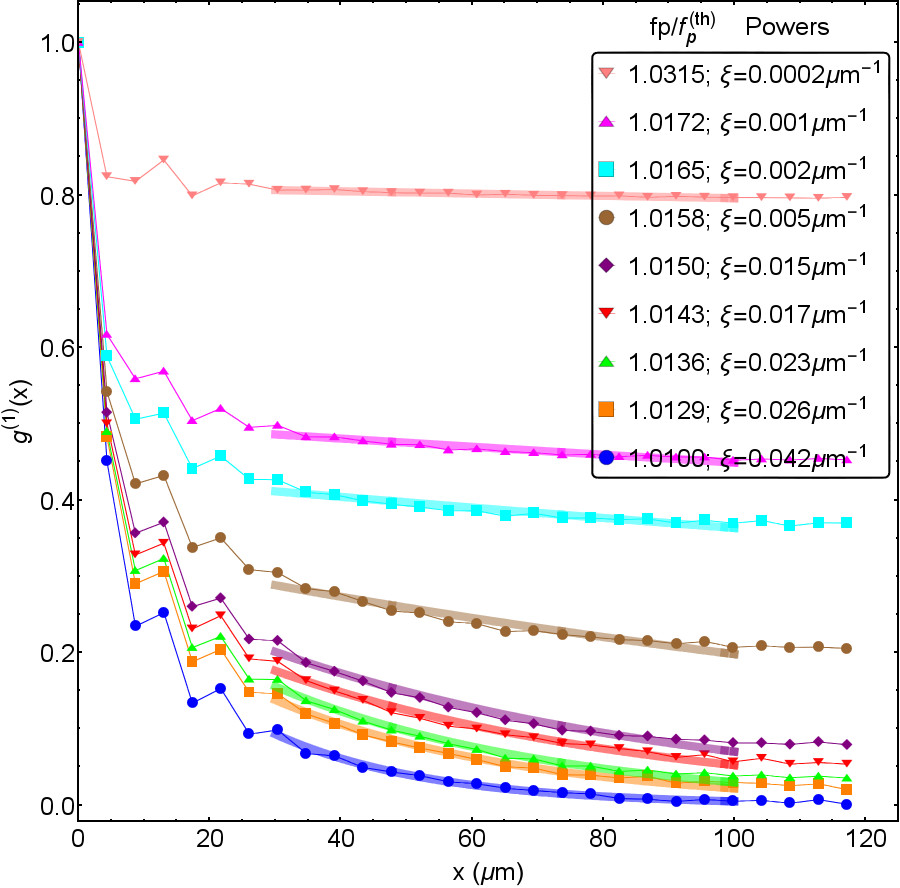}
\includegraphics[width=0.49\linewidth]{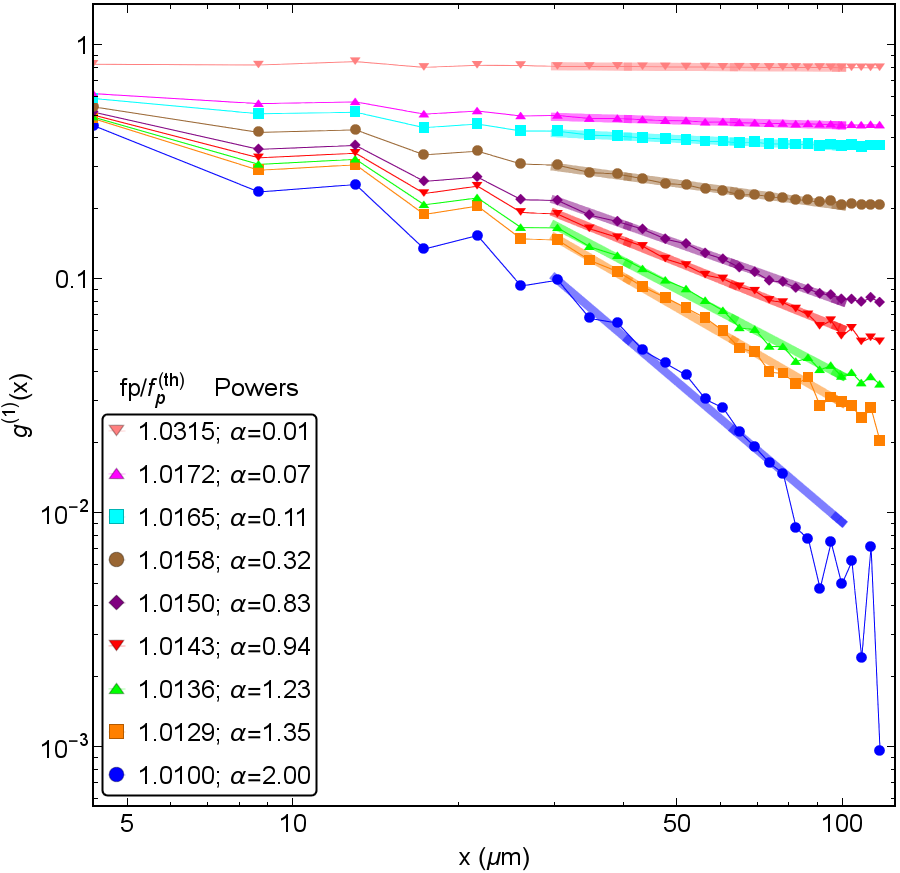}
\includegraphics[width=0.49\linewidth]{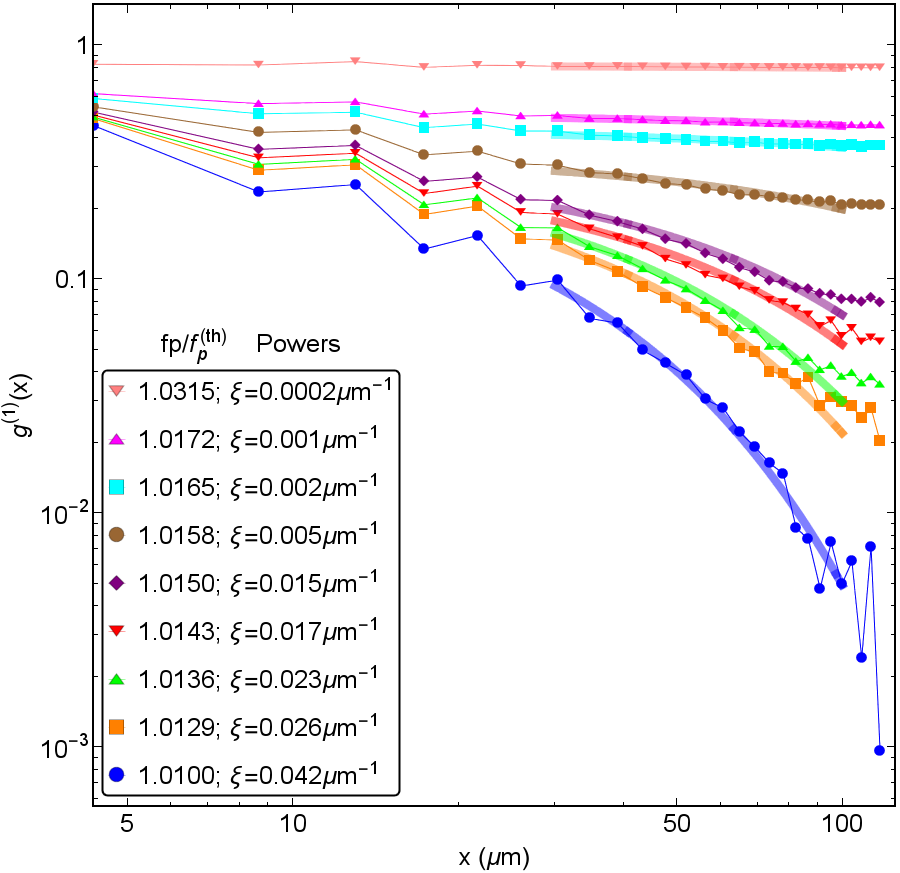}
\caption{ {\bf Algebraic and exponential fits to the
    \mbox{correlation} function.}
Here we show both the algebraic (left) and exponential (right) fits to
the correlation function on a full linear scale (above) and on an
exponential scale (below).
The fitting to the data is taken over the range
$30\mu\textrm{m}<x<100\mu\textrm{m}$. This domain needs to avoid both
the region close to origin with short-range correlations and the
edges, where the behaviour might be influenced by periodic boundary
conditions.  }
\label{fig:g1_linlin}
\end{figure*}

The behaviour of the first order correlation function $g^{(1)}(x)$,
whether it decays as an exponential or as a power law, is determined
by fitting the tail of the data to each functional form and taking the
better fit of the two. The fitting is illustrated in
Fig. \ref{fig:g1_linlin}.
For pump powers close but above the critical region it is clear that
the power-law fits the data better (which is also confirmed by a
larger correlation coefficient), while below the critical region the
exponential fit is closer to the data.
For stronger pump powers, away from the transition, $g^{(1)}(x)$ is
practically constant on these length-scales (as reported in
experiments and discussed in the main text).

\end{document}